\documentstyle[12pt]{article}

\begin{document}

\begin{titlepage}

\hoffset = .5truecm
\voffset = -2truecm

\centering

\null

{\normalsize \sf \bf International Atomic Energy Agency\\
and\\
United Nations Educational, Scientific and Cultural Organization\\}
\vskip 1truecm
{\huge \bf
INTERNATIONAL CENTRE\\
FOR\\
THEORETICAL PHYSICS\\}
\vskip 3truecm

{\LARGE \bf
Jacobian Elliptic Wave Solutions for the Wadati-Segur-Ablowitz Equation
}\\
\vskip 1truecm

{\large \bf
Chooi-Gim Rosy Teh,
\\}
\medskip
{\large \bf
W. K. Koo, 
\\}
\medskip
{\large and\\}
\medskip
{\large \bf
B. S. Lee. 
\\}

\vskip 8truecm

{\bf MIRAMARE--TRIESTE\\}
July 1996

\end{titlepage}

\hoffset = -1truecm
\voffset = -2truecm

\title{\bf
Jacobian Elliptic Wave Solutions for the Wadati-Segur-Ablowitz Equation}

\author{
{\bf Chooi-Gim Rosy Teh}\\
{\normalsize International Centre for Theoretical Physics, Trieste 34100,}
{\bf Italy}\\
{\normalsize and}\\
{\normalsize ITM/ADP, Centre for Preparatory Education, Sect. 17, 
40200 Shah Alam,} 
{\bf Malaysia}\thanks{Permanent Address}\\ 
{\bf W. K. Koo}\\
{\normalsize Kolej Damansara Utama, Penang, Jalan Padang Kota Lama, 
10200 Penang,}
{\bf Malaysia}\\
{\bf B. S. Lee}\\
{\normalsize School of Physics, University of Science of Malaysia, 
11800 USM Penang,}
{\bf Malaysia}}

\date{15th July 1996}
\newpage

\maketitle

\begin{abstract}
Jacobian elliptic travelling wave solutions for a 
new Hamiltonian amplitude equation determining some instabilities of 
modulated wave train are  
obtained. By mere variation of the Jacobian elliptic parameter $k^2$ from 
zero to one, these solutions are transformed from a trivial one to the known  
solitary wave solutions \cite{kn:1}, \cite{kn:2}.
\end{abstract}

\newpage



\section{Introduction}

A new Hamiltonian amplitude equation governing modulated wave 
instabilities was reported by Wadati et. al 
\cite{kn:1} in 1992. This 
new Hamiltonian amplitude equation that is introduced, 

\begin{equation}
i\psi_x + \psi_{tt} + 2\sigma|\psi |^2\psi - \epsilon\psi_{xt} = 0,~~
\sigma = \pm 1,~~ \epsilon\ll 1,
\label{eq:1}
\end{equation}

\noindent governs certain instabilities of modulated wave trains. 
The subscripts 
here denote partial derivatives. The addition of the term $-\epsilon 
\psi_{xt}$ overcomes the ill-posedness of the unstable nonlinear 
Schrodinger equation. Wadati et. al  
\cite{kn:1} found that this new equation is
apparently not integrable because it does not satisfy the Painleve 
property but it is a Hamiltonian analogue of the 
Kuramoto-Sivashinsky equation which arises in dissipative systems.

By assuming a boundary condition for $\psi (x, t)$ such that it is either 
rapidly decreasing or periodic in $x$, Eq. (\ref{eq:1}) has at least 
three conserved 
quantities given by \cite{kn:1}

\begin{eqnarray}
I_1 = \int dx (-i\psi^*_t\psi + i\psi^*\psi_t + \frac{1}{2}i\epsilon 
\psi^*_x\psi - \frac{1}{2}i\epsilon\psi^*\psi_x ),
\label{eq:2}\\
I_2 = \int dx (\frac{i}{2}\psi^*\psi_x - \frac{i}{2}\psi^*_x\psi +
\sigma |\psi|^4 + \psi^*_t\psi_t),
\label{eq:3}
\end{eqnarray}

which is the Hamiltonian and

\begin{equation}
I_3 = \int dx (\psi^*_x\psi_t + \psi^*_t\psi_x - \epsilon\psi^*_x\psi_x).
\label{eq:4}
\end{equation}

The two solitary wave solutions that have been reported are \cite{kn:1}

\begin{equation}
\psi (x, t) = \pm\sqrt{ - \frac{2m}{n}} sech \left( \sqrt {- \frac{m}{l}}
(x - vt + \xi_0)\right) e^{i(Kx - \Omega t)},
\label{eq:5}
\end{equation}

\noindent where $\sigma = 1,~~l = v^2 + \epsilon v > 0,~~-m = K + \Omega^2 + 
\epsilon K \Omega > 0,~~n = 2\sigma,~~\xi_0 =$ constant, and \cite{kn:2}

\begin{equation}
\psi (x, t) = \pm\sqrt{ - \frac{m}{n}} tanh \left( \sqrt {\frac{m}{2l}}
(x - vt + \xi_0)\right) e^{i(Kx - \Omega t)},
\label{eq:6}
\end{equation}

\noindent where $\sigma = -1,~~l = v^2 + \epsilon v > 0,~~
-m = K + \Omega^2 + \epsilon K\Omega < 0$.

In this paper, we show that Eq. (\ref{eq:1}) does not only support the 
solitary wave solutions (\ref{eq:5}) and (\ref{eq:6}) but it also 
supports Jacobian 
elliptic travelling wave solutions as well. 
The solitary wave solutions (\ref{eq:5}) and (\ref{eq:6})
can be obtained from the Jacobian elliptic wave solutions when the 
Jacobian elliptic parameter $k^2$ takes the value one for the three basic 
Jacobian elliptic functions, $sn(u), cn(u),$ and $dn(u)$. These elliptic wave
solutions will be discussed in the next section and we will end with some 
remarks in the last section.

\newpage

\section{The Jacobian Elliptic Wave Solutions}

We start off with the travelling wave ansatz of Wadati et. al \cite{kn:1}, 

\begin{equation}
\psi (x, t) = e^{i\eta} \phi (\xi),~~~\xi = x - vt,~~~\eta = Kx - \Omega t,
\label{eq:7}
\end{equation}

\noindent and find that 

\begin{eqnarray}
\psi_x = e^{i\eta}\left(\phi^\prime (\xi) + iK\phi(\xi)\right),\nonumber\\
\psi_{xt} = e^{i\eta} \left(-v\phi^{\prime\prime} (\xi) - i(Kv + \Omega)
\phi^{\prime}(\xi) + K\Omega\phi(\xi)\right),\nonumber\\
\psi_{tt} = e^{i\eta} \left(v^2 \phi^{\prime\prime} (\xi) + 2i\Omega v
\phi^{\prime}(\xi) - \Omega^2\phi(\xi)\right).
\label{eq:8}
\end{eqnarray}
Upon substitution of Eqs. (\ref{eq:8}) into Eq. (\ref{eq:1}), 
the Hamiltonian amplitude equation then reads,

\begin{eqnarray}
(v^2 + \epsilon v)\phi^{\prime\prime}(\xi) + i[1 + 2\Omega v + \epsilon
(Kv + \Omega)]\phi^{\prime} (\xi)\nonumber\\ 
- (K + \Omega^2 + \epsilon K\Omega)\phi (\xi) + 2\sigma\phi^3(\xi) = 0.
\label{eq:9}
\end{eqnarray}
As exact, explicit solutions cannot be found for Eq. (\ref{eq:9}) without 
making,
$$ 1 + 2\Omega v + \epsilon(Kv + \Omega) = 0, $$
as had been done by Wadati et. al \cite{kn:1}, we choose to do the same
and reduce Eq. (\ref{eq:9}) into the solvable form,

\begin{eqnarray}
l\phi^{\prime\prime}(\xi) + m\phi(\xi) + n\phi^3(\xi) = 0,\nonumber\\
l = v^2 + \epsilon v,~~~\Omega = -\left(\frac{1 + \epsilon vK}
{2v + \epsilon}\right),\nonumber\\
m = -(K + \Omega^2 + \epsilon K\Omega)\nonumber\\
= - \frac{ \left((1 + 4Kv^2) + \epsilon(4Kv) - \epsilon^2(K^2 v^2) - 
\epsilon^3(K^2 v)\right)}{(2v + \epsilon)^2},\nonumber\\
n = 2\sigma.
\label{eq:10}
\end{eqnarray}

By noticing that Eq. (\ref{eq:10}) is just the Jacobian elliptic functions 
differential equation \cite{kn:3} - \cite{kn:5},
the solution to Eq. (\ref{eq:10}) is therefore given by

\begin{equation}
\phi (\xi) = f E(u);~~~~~u = g(\xi + \xi_0),
\label{eq:11}
\end{equation}
where $ f,~~g$, and $\xi_0$ are constants, and $E(u)$ is any of the twelve 
Jacobian elliptic functions.
Upon substitution, Eq. (\ref{eq:10}) reduces to the Jacobian 
elliptic functions differential equation, 

\begin{eqnarray}
E^{\prime\prime}(u) + a E(u) + b E^3(u) = 0,\nonumber\\
(E^\prime(u))^2 + a E^2(u) + \frac{1}{2} b E^4(u) = c,
\label{eq:12}
\end{eqnarray}
where $ a = a(k) = \frac{m}{lg^2},~~~b = b(k) = \frac{nf^2}{lg^2},~~~c = $
integration constant, and $ 0\leq k^2 \leq 1$ is the Jacobian elliptic
parameter. The constants of solution (\ref{eq:11}) can also be written as

\begin{eqnarray}
g = \sqrt{\frac{m}{la}} = \sqrt{\frac{-(1 + 4Kv^2) - \epsilon(4Kv) +
\epsilon^2(K^2 v^2) + \epsilon^3(K^2 v)}{a(2v + \epsilon)^2 (v^2 + \epsilon v)
}},\nonumber\\
f = \sqrt{\frac{mb}{na}} = \sqrt{\frac{b[-(1 + 4Kv^2) - \epsilon(4Kv) +
\epsilon^2(K^2 v^2) + \epsilon^3(K^2 v)]}{2\sigma a(2v + \epsilon)^2}},  
\label{eq:13}
\end{eqnarray}
and $\xi_0$ is an arbitrary constant. The constants $ a, b$, and $c$ for the 
twelve Jacobian elliptic functions are, \cite{kn:3} 

$$
\begin{array}{llllllll}
E(u)       & a           & b            & c            & 0\leq k^2\leq 1\\
sn(u)      & 1+k^2       & -2k^2        & 1            & 0\leq k^2\leq 1\\
cn(u)      & 1-2k^2      &  2k^2        & 1-k^2        & k^2\neq\frac{1}{2} \\
dn(u)      & -(2-k^2)    &  2           & -(1-k^2)     & 0\leq k^2\leq 1\\
ns(u)      & 1+k^2       & -2           & k^2          & 0\leq k^2\leq 1\\
nc(u)      & 1-2k^2      & -2(1-k^2)    & -k^2         & k^2\neq\frac{1}{2} \\  
nd(u)      & -(2-k^2)    & 2(1-k^2)     & -1           & 0\leq k^2\leq 1\\  
sc(u)      & -(2-k^2)    & -2(1-k^2)    & 1            & 0\leq k^2\leq 1\\                   
sd(u)      & 1-2k^2      & 2k^2(1-k^2)  & 1            & k^2\neq\frac{1}{2} \\
cs(u)      & -(2-k^2)    & -2           & 1-k^2        & 0\leq k^2\leq 1\\
cd(u)      & 1+k^2       & -2k^2        & 1            & 0\leq k^2\leq 1\\
ds(u)      & 1-2k^2      & -2           & -k^2(1-k^2)  & k^2\neq\frac{1}{2} \\ 
dc(u)      & 1+k^2       & -2           & k^2          & 0\leq k^2\leq 1\\
\end{array}
$$

Since the wave solution, $\phi(\xi)$, is singular when $a=0$, the parameter
$k^2$ must not take on the value $\frac{1}{2}$ when $E(u)$ is $cn(u),~~
nc(u),~~sd(u),$ or $ds(u)$. However for the other eight Jacobian elliptic 
solutions, $a\neq 0$ when $k^2$ runs from zero to one.

We note that the Jacobian elliptic solution, 

\begin{equation}
\psi (x, t) = \pm\sqrt{ - \frac{2m k^2}{n(1+k^2)}} sn \left( 
\sqrt {\frac{m}{l(1+k^2)}}
(\xi + \xi_0)\right) e^{i(Kx - \Omega t)},
\label{eq:14}
\end{equation}
where $\frac{m}{n}<0,~~\frac{m}{l}>0$; or equivalently
$\sigma =-1,~~l = v^2 + \epsilon v > 0,~~-m = K + \Omega^2 + 
\epsilon K \Omega < 0$, reduces to the solitary wave solution of Kong
and Zhang \cite{kn:2} when the parameter $k^2 = 1$, that is,
$$
\phi (\xi) = \pm\sqrt{ - \frac{m}{n}} tanh \left( \sqrt {- \frac{m}{2l}}
(\xi + \xi_0)\right).
$$
However $\phi(\xi)$ tends to zero as $k^2\rightarrow 0$. Hence the 
solution (\ref{eq:14}) evolves from the trival zero solution to the solitary 
wave solution of Eq. (\ref{eq:6}), as $k^2$ runs from zero to one.

Similarly the Jacobian elliptic wave solutions,

\begin{equation}
\psi (x, t) = \pm\sqrt{-\frac{2m}{n(2-k^2)}} dn \left( \sqrt {- 
\frac{m}{l(2-k^2)}}(\xi + \xi_0)\right) e^{i(Kx - \Omega t)},
\label{eq:15}
\end{equation}
with $0\leq k^2\leq 1$ and

\begin{equation}
\psi (x, t) = \pm\sqrt{\frac{2mk^2}{n(1-2k^2)}} cn \left( 
\sqrt {\frac{m}{l(1-2k^2)}}(\xi + \xi_0)\right) e^{i(Kx - \Omega t)},
\label{eq:16}
\end{equation}
with $\frac{1}{2}< k^2\leq 1$,  
reduce to the solitary wave solution of Wadati 
et. al \cite{kn:1} when $k^2=1$, that is,

$$
\phi (\xi) = \pm\sqrt{ - \frac{2m}{n}} sech \left( \sqrt {- \frac{m}{l}}
(\xi + \xi_0)\right).
$$
Here we have $\frac{m}{n}<0,~~\frac{m}{l}<0$; or equivalently
$\sigma =+1,~~l = v^2 + \epsilon v > 0,~~-m = K + \Omega^2 + 
\epsilon K \Omega > 0$, for both the solutions (\ref{eq:15}) and  
(\ref{eq:16}). Solution (\ref{eq:15}) unlike soluton (\ref{eq:16}), 
approaches the trivial solution,

\begin{equation}
\psi (x, t) = \pm\sqrt{-\frac{m}{n}} e^{i(Kx - \Omega t)}.
\label{eq:17}
\end{equation}
when $k^2$ approaches zero,
whereas solution (\ref{eq:16}) blows up when $k^2$ approaches $\frac{1}{2}$, 
as it is singular there at $k^2=\frac{1}{2}$.

Hence we see that the solitary wave solution (\ref{eq:5}), unlike the 
solitary wave solution (\ref{eq:6}), actually evolves from two solutions 
(\ref{eq:15}) and (\ref{eq:16}), instead of one. This shows that bifurcation 
in the conserved quantities, Eq. (\ref{eq:2}) to Eq. (\ref{eq:4}), may occur 
here. Further work in this area is still in progress.

\section{Remarks}

1) The nonlinear Schrodinger equation,

\begin{equation}
i\psi_x + \psi_{tt} + 2\sigma|\psi |^2\psi = 0,~~\sigma = \pm 1,
\label{eq:18}
\end{equation}
which has been proposed as a model for nonlinear modulation of stable plane
waves in unstable media \cite{kn:6} - \cite{kn:9} is a special case of the 
new Hamiltonian amplitude equation (\ref{eq:1}), when $\epsilon = 0$ 
\cite{kn:1}. Hence beside the two 
solitary wave solutions of Ref. \cite{kn:1} and \cite{kn:2}, 
the nonlinear Schrodinger 
equation (\ref{eq:18}) also possesses Jacobian elliptic travelling 
waves solutions,

\begin{equation}
\psi(x, t) = \pm\frac{1}{2v}\sqrt{-\frac{b(1+4Kv^2)}{2a\sigma}} 
E\left(\frac{1}{2v^2}\sqrt{-\frac{(1+4Kv^2)}{a}}(x-vt)\right)
e^{i(Kx-\Omega t)}
\label{eq:19}
\end{equation}
when $E(u)$ is any of the twelve Jacobian elliptic functions.

2) The conserved quantities of Eq. (\ref{eq:2}) to (\ref{eq:4}) when
integrated over all space is finite only for the solitary wave solution
(\ref{eq:5}), and not the solitary wave solution (\ref{eq:6}). 
However when these conserved quantities  are integrated over half a period,
that is from $-{\bf K}(k)$ to ${\bf K}(k)$ \cite{kn:3} - \cite{kn:5}, then
they are finite only for the solutions when $E(u)$ is $cnu,$ $snu,$ 
$dnu,$ $ndu,$ $sdu,$ and $cdu$ as these solutions are regular over all space.
More will be discussed about these conserved quantities in a longer 
report.

3) Beside the solutions when $E(u)$ is $nsu,~ncu,~scu,~csu,~dcu,$ 
and $dsu$ which are singular, all the other Jacobian elliptic wave 
solutions are regular over all space with boundary condition that is 
either rapidly decreasing or periodic in $x$. We believe that these
Jacobian elliptic wave solutions had been overlooked by Wadati et. al
\cite{kn:1} and Kong and Zhang \cite{kn:2} as these authors
were not looking for periodic solutions but rather solutions with 
boundary condition that is rapidly decreasing in $x$.

4) These Jacobian elliptic wave solutions are the most general solutions
of the Wadati-Segur-Ablowitz equation;   
all the other solitary wave solutions arise from changing the parameter
$k^2$, or correspondingly in an experimental setup the experimental 
condition, which we can control sometimes.

\section{Acknowledgements}

The author, Chooi-Gim Rosy Teh, would like to thank 
Prof. Seifallah Randjbar-Daemi for reading through the manuscript,
the Bureau of Research and Consultancy Centre, ITM for a short term 
grant, and the Japanese Government for financial support. 
She would also like to acknowledge the hospitality of the 
International Centre of Theoretical 
Physics, and the leave of absence from her home institute, Institute of
Technology of MARA, for which this work would not have been possible. 

The authors W. K. Koo and B. S. Lee would also like to thank University 
of Science of Malaysia for a short term grant.

\end{document}